\documentclass[12pt,letterpaper]{article}
\usepackage{amssymb,amsmath,amsthm,enumerate,nicefrac}
\usepackage{graphicx, color}
    \usepackage[driverfallback=hypertex,pagebackref=true,colorlinks]{hyperref}
    \hypersetup{linkcolor=[rgb]{.7,0,0}}
    \hypersetup{citecolor=[rgb]{0,.7,0}}
    \hypersetup{urlcolor=[rgb]{.7,0,.7}}
\usepackage{geometry}
\usepackage{epstopdf}
\usepackage[ruled]{algorithm2e}
\usepackage{braket}

\AtBeginDocument{%
  \addtolength\abovedisplayskip{-0.25\baselineskip}%
  \addtolength\belowdisplayskip{-0.15\baselineskip}%
}

\usepackage{float}
\floatstyle{ruled}
\newfloat{algorithm}{tbp}{loa}
\providecommand{\algorithmname}{Algorithm}
\floatname{algorithm}{\protect\algorithmname}

\usepackage[titletoc,title]{appendix}
\usepackage{cleveref,url}

\makeatletter
\newtheorem*{rep@theorem}{\rep@title}
\newcommand{\newreptheorem}[2]{%
\newenvironment{rep#1}[1]{%
 \def\rep@title{#2 \ref{##1}}%
 \begin{rep@theorem}}%
 {\end{rep@theorem}}}
\makeatother

\newtheorem{lemma}{Lemma}[section]
\newtheorem{claim}{Claim}[section]

\newtheorem{definition}[lemma]{Definition}

\newtheorem{proposition}[lemma]{Proposition}
\newreptheorem{proposition}{Proposition}

\def\Nbb{{\mathbb {N}}}
\newcommand{\eps}{\varepsilon}
\def\Rbb{{\mathbb {R}}}

\def\poly{{\rm {poly}}}

\def\V{{\mathcal {V}}}

\def\supp{{\rm supp}}

\DeclareMathOperator*{\p}{\Pr}

\renewcommand{\eps}{\varepsilon}

\newcommand{\abs}[1]{\left| #1 \right|}
\newcommand{\vabs}[1]{\left\| #1 \right\|}

\newcommand{\pbra}[1]{\left( #1 \right)}
\newcommand{\sbra}[1]{\left[ #1 \right]}

\newcommand{\BPL}{\mathsf{BPL}}

\newcommand{\LSPACE}{\mathsf{L}}
\newcommand{\BQL}{\mathsf{BQL}}

\title{Quantum Logspace Computations are Verifiable}
\date{}
\author{Uma Girish\thanks{Princeton University. E-mail: ugirish@cs.princeton.edu. Research supported by a Simons Investigator Award, by the National Science Foundation grants No. CCF-1714779, CCF-2007462
and by the IBM Phd Fellowship.}
\and Ran Raz\thanks{Princeton University. E-mail: ranr@cs.princeton.edu. Research supported by a Simons Investigator Award and by the National Science Foundation grants No. CCF-1714779, CCF-2007462.}
\and Wei Zhan\thanks{Princeton University. E-mail: weizhan@cs.princeton.edu. Research supported by a Simons Investigator Award and by the National Science Foundation grants No. CCF-1714779, CCF-2007462.}}

\date{}
\begin{document}
\maketitle

\begin{abstract} 
In this note, we observe that quantum logspace computations are verifiable by classical logspace algorithms, with unconditional security.
More precisely, every language in $\BQL$ has an (information-theoretically secure) streaming proof with a quantum logspace prover and a classical logspace verifier. The prover provides a polynomial-length proof that is streamed to the verifier. The verifier has a read-once one-way access to that proof and 
is able to verify that the computation was performed correctly. That is, if the input is in the language and the prover is honest, the verifier accepts with high probability, and, if the input is not in the language, the verifier rejects with high probability even if the prover is adversarial.
Moreover, the verifier uses only $O(\log n)$ random
bits.


\end{abstract}

\section{Introduction} 

The problem of how to classically verify that a quantum computation was performed correctly, first suggested by Gottesman in 2004, has been studied in numerous recent works (see for example~\cite{bfk,ruv,fk,abem,Mahadev,cbgjv,ccy,acgh,bklmmvvy}). Mahadev's breakthrough work presented the first protocol for classical verification of quantum computations~\cite{Mahadev}. Her protocol is only secure against computationally bounded adversarial provers, under cryptographic assumptions. In this note we observe that for quantum logspace computations, there is a simple verification protocol, with a classical logspace verifier, such that the protocol is secure against adversarial provers with unlimited computational power. Moreover, the protocol is non-interactive.
Our proof is similar to our recent proof that shows that randomized logspace computations are verifiable using
only $O(\log n)$ random bits~\cite{grz23}.

\subsection{Streaming Proofs}

A streaming proof consists of a pair of (classical or quantum) randomized algorithms, a prover and a verifier, which share a common stream tape. In our work,  the prover is a quantum logspace machine and the verifier is a classical randomized logspace machine. The prover doesn't have a separate  output tape, instead, it has write-once access to the proof tape onto which it writes a classical bit string $\Pi$. The verifier has read-once access to the proof tape from which it can read $\Pi$. Both the verifier and the prover have read-many access to the input $x\in \{0,1\}^*$. We allow the prover and verifier to output a special symbol $\bot$. Upon outputting this symbol, the algorithm stops all further processing and we say that the algorithm aborts.

\begin{definition}[Logspace Streaming Proofs] Let $\mathcal{F}=\{f_n: \{0,1\}^n\rightarrow \{0,1\}\}_{n\in \Nbb}$ be a family of functions. Let $P:\Nbb\to\Nbb$ be a monotone computable function. We say that $\mathcal{F}$ has a logspace streaming proof of length $P$ if there is a (possibly quantum) logspace prover $\mathcal{P}$ and a classical randomized logspace verifier $\V$, that uses a random string $R$, such that on input $x\in \mathrm{supp}(f_n)$,
	\begin{enumerate}\item The honest prover $\mathcal{P}$, with at least $\tfrac{3}{4}$ probability, outputs a (randomized) proof $\Pi\in\{0,1\}^{P(n)}$ such that \[\underset{R}{\Pr}[\V(x,\Pi)=f_n(x)]\ge \tfrac{3}{4}\]
		(where the probability is over the uniform distribution over $R$.)
		\item For an arbitrary $\Pi\in\{0,1\}^{P(n)}$ (even adversarially chosen after seeing the input $x$),  \[\underset{R}{\Pr}[\V(x,\Pi)\in\{f_n(x),\bot\}] \ge \tfrac{3}{4}\]
		(where the probability is over the uniform distribution over $R$.)
	\end{enumerate}
	Let $k:\Nbb\to \Nbb$ be a monotone computable function. If the verifier  $\V$ never reads more than $k(n)$ random bits from $R$, we say that the verifier uses at most $k(n)$ random bits. 
	
\end{definition}

We sometimes omit the length of the proof and it is understood that $P$ is at most the runtime of the prover, which is polynomial in $n$. 

\subsection{Our Main Result}

Our main result is as follows.

\begin{proposition}\label{maintheorem2}
A language is in $\BQL$ if and only if it has a streaming proof between a quantum logspace prover and a classical logspace verifier where the verifier uses $O(\log n)$ random bits.
\end{proposition}

\section{Preliminaries}

Let $n\in\Nbb$. We use $[n]$ to denote $\{1,2,\ldots,n\}$. Let $v\in \Rbb^n$. For $i\in[n]$ we use $v_i$ to denote the $i$-th coordinate of $v$. Let $1\le k<\infty$. Let $\|v\|_k:=\pbra{\sum_{i\in [n]} \abs{v_i}^k}^{1/k}$ denote the $\ell_k$-norm of~$v$. This induces an operator norm on matrices $M\in \Rbb^{n\times n}$ by $\|M\|_k:=\max_{v\in \Rbb^n\setminus\{ \vec{0}\}}\tfrac{ \|M(v)\|_k}{\|v\|_k}$. This norm is sub-multiplicative, i.e., $\|M\cdot N\|_k\le \|M\|_k\cdot \|N\|_k$ for all $M,N\in \Rbb^{n\times n}$. Let $\|v\|_\infty =\max_{i\in [n]}\abs{v_i}$ denote the $\ell_\infty$-norm of $v$ and let $\|M\|_{\max}=\max_{i,j\in[n]}\abs{M_{i,j}}$ (this is not an induced operator norm). We have the following inequalities for all $M\in \Rbb^{n\times n}, v\in \Rbb^n$ and $1\le k,k'<\infty $. 
\[ \|M\|_{\max} \le \|M\|_k \le n\cdot \|M\|_{\max} \]
\[k\ge k'\implies \|v\|_k \le \|v\|_{k'} \]
We use $M[i,j]$ to refer to the $(i,j)^{\text{th}}$ entry of the matrix $M$.

\subsection{Our Model of Computation}

In this work, a deterministic Turing machine consists of a read-only input tape, a work tape and a write-once output tape. A randomized Turing Machine has an additional read-once randomness tape consisting of random bits. Let $S,T,R:\Nbb\rightarrow\Nbb$ be any monotone computable functions. We typically use $S$ to denote the space complexity and $T$ to denote the time complexity of a Turing Machine. When we say that an event occurs with high probability, we typically mean that it occurs with probability at least $2/3$. An algorithm is said to have bounded error if the probability of error is at most 1/3. By standard error-reduction techniques, we could choose this number to be any constant in $(0,1/2)$.

A deterministic (resp.  bounded-error randomized) $(S,T)$ algorithm refers to a deterministic (resp. randomized) Turing Machine such that for all $x\in \{0,1\}^*, |x|=n,$ the machine with $x$ on its input tape, uses at most $S(n)$ bits of space on its work tape and runs in at most $T(n)$ time. We say that an algorithm computes a family of functions $\{f_n: \{0,1\}^n\rightarrow \{0,1\}^*\}_{n\in \Nbb}$ if for all $n\in \Nbb,x\in\{0,1\}^n$, the output of the algorithm on input $x$ is $f_n(x)$ (with high probability if the algorithm is bounded-error randomized).  These functions may be partial, i.e., defined on a strict subset of $\{0,1\}^n$. The Turing machine is said to use $R$ bits of randomness if on inputs of size $n\in\Nbb$, the machine never reads more than $R(n)$ bits on the randomness tape.

\paragraph*{Logspace Computation:}

 A logspace algorithm refers to an $(O(\log( n)),\poly(n))$ algorithm. The (promise) class $\LSPACE$ refers to all families of single-bit-output functions computable by deterministic logspace algorithms. The (promise) class $\BPL$ refers to all families of single-bit-output functions computable by randomized logspace algorithms with high probability. All these classes are inherently promise classes, so for the rest of the paper, we omit this prefix. We use the notation family of functions, languages and problems interchangeably.  It is possible to define quantum analogues of the aforementioned complexity classes. In particular, we will be interested in $\BQL$, the set of all families of single-bit-output functions computable by quantum logspace algorithms. The readers are referred to \cite{fr20} for a formal definition of $\BQL$, while here we characterize the class using a complete problem stated below.

\subsection{Unitary Matrix Powering}
We consider the following promise problem.

\begin{definition}[Unitary Matrix Powering]
	The inputs are an $n\times n$ unitary matrix $M$ and a parameter $T\le \poly(n)$ and an $n\times n$ projection matrix $\Pi$ onto a subset of standard basis states. The promise on the input is that $\|\Pi M^T(e_1)\|_2^2\ge 4/5$ or $\|\Pi M^T(e_1)\|_2^2\le 1/5$. The goal is to output 1 in the former case and 0 in the latter case.
\end{definition}

\begin{proposition}
	The Unitary Matrix Powering Problem is logspace-complete for $\BQL$.\label{bqlcomplete}
\end{proposition} 

The proof of this is deferred to the appendix.

\section{Classical Logspace Verifiers for Quantum Logspace Computations}

In this section, we prove \Cref{maintheorem2} which states that a family of functions is in $\BQL$ if and only if it has a logspace streaming proof between a quantum prover and a classical verifier that reads $O(\log n)$ random bits.  First, it is clear that any streaming proof between a quantum logspace prover and a classical logspace verifier can be implemented by a $\BQL$ algorithm, which simulates the hoest prover and the verifier, with success probability at least $(3/4)^2>1/2$ which can be amplified. It suffices to argue that the Unitary Matrix Powering Problem can be solved by a streaming proof between a quantum logspace prover and a classical logspace verifier, where the verifier uses $O(\log n)$ random bits. Towards this, we define a notion of a $\delta$-good sequence of vectors for a matrix $M$. 

\begin{definition}
	Let $M$ be any $n\times n$ matrix and $T\le \poly(n)$ be a natural number. Let $v_i= M^i(e_1)$ for all $i\le T$. Let $\delta\in[0,1]$. A sequence of vectors $v'_0,v'_1,\ldots,v'_T\in \mathbb{R}^n$ is said to be $\delta$-good for $M$ if for all $i\in [T]$, we have $\|v'_i-v_i\|_2 \le \delta$ and $v_0=e_1$. 
\end{definition}

We make use of the following claims.

\begin{claim} There is a quantum logspace prover which given an $n\times n$ unitary matrix $M$ and parameters $T\le \poly(n),\delta\ge\frac{1}{\poly(n)}$ as input, outputs a $\delta$-good sequence of vectors for $M$ with probability at least $\tfrac{3}{4}$.  \label{claimqprover} \end{claim}

\begin{claim} \label{claimverifier} Let $\tfrac{1}{\poly(n)}<\delta \le \tfrac{1}{10^4  T^2}$.  There is a randomized logspace verifier which given any $n\times n$ unitary matrix $M$ and parameters $T\le \poly(n),\delta$ as input and read-once access to a stream of vectors $v'_0,\ldots,v'_T\in \mathbb{R}^n$ (where each vector is specified up to $\Theta(\log (n))$ bits of precision), does the following.
	\begin{itemize}
		\item If the sequence is $\delta$-good for $M$, then the probability that the algorithm aborts is at most $1/4$. 
		\item If $\vabs{v'_T-v_T}_2 \ge   \tfrac{1}{5}$, then the algorithm aborts with probability at least $3/4$. 
	\end{itemize} 
	Furthermore, this algorithm only uses $O(\log (n))$ bits of randomness. 
\end{claim}

Let us see how to complete the proof using \Cref{claimqprover} and \Cref{claimverifier}. Given an $n\times n$ unitary matrix $M$ as input and a parameter $T\le \poly(n)$, set $\delta= \min\left\{\tfrac{1}{10^4 T^2},\tfrac{1}{10}\right\}$. Run the prover's algorithm from \Cref{claimqprover} using this value of $\delta$ to produce a stream $v'_0,\ldots,v'_T$. Run the verifier's algorithm from \Cref{claimverifier} on this stream to verify. If it doesn't abort, we have the verifier return $1$ if $\vabs{\Pi( v'_T)}_2^2\ge 0.6$, return $0$ if $\vabs{\Pi (v'_T)}_2^2\le 0.4$ and return $\bot$ otherwise. With the access to read $\Pi$ from the input, this computation can be easily done in classical logspace when $v'_T$ is given as a stream.

\paragraph*{Completeness:} \Cref{claimqprover} implies that an honest prover outputs a $\delta$-good sequence with probability at least $\tfrac{3}{4}$. \Cref{claimverifier} implies that an honest proof is aborted with probability at most $\tfrac{1}{4}$. Since $\vabs{v'_T-v_T}_2  \le \delta\leq 1/10$ by assumption and $\Pi$ is a projection, $\vabs{\Pi(v'_T)-\Pi(v_T)}_2\le 1/10$. Hence, if $\vabs{\Pi ({v}_T)}_2^2\ge 4/5$, then $\vabs{\Pi (v'_T)}_2^2\ge (\sqrt{4/5}-0.1)^2\ge 0.6$ and if $\vabs{\Pi (v_T)}_2^2\le 1/5$ then $\vabs{\Pi (v'_T)}_2^2 \le (\sqrt{1/5}+0.1)^2\le 0.4$. Thus, the verifier will return the correct answer whenever the sub-routine doesn't abort.

\paragraph*{Soundness:} Consider the behavior of this verifier on an arbitrary proof. If the verifier makes a mistake and returns the incorrect answer, it must be the case that either $\vabs{\Pi ({v}_T)}_2^2\ge 4/5$ and $\vabs{\Pi (v'_T)}_2^2\le 0.4$ or $\vabs{\Pi ({v}_T)}_2^2\le 1/5$ and $\vabs{\Pi (v'_T)}_2^2\ge 0.6$. In either case, we must have $\vabs{v'_T-v_T}_2 \ge \min\pbra{ \sqrt{4/5}-\sqrt{0.4},\sqrt{0.6}-\sqrt{1/5} }\ge \tfrac{1}{5}$. \Cref{claimverifier} implies that such a proof is aborted with probability at least $\tfrac{3}{4}$. This completes the proof of \Cref{maintheorem2}. 

We now proceed to prove \Cref{claimqprover}. 

\begin{proof}[Proof of \Cref{claimqprover}]The prover starts by outputting $v_0=e_1$. To output the intermediate $v_i$, we make use of the following result from~\cite{grz20}. It appears as Corollary~15 and we paraphrase it as follows.
\begin{lemma}\label{grzlemma}
Given an $n\times n$ matrix $M$ with $\vabs{M}_2\le 1$, a positive integer $i\le \poly(n)$, two unit vectors $v,w\in \mathbb{R}^n$ and an error parameter $\delta>0$, there is a quantum algorithm with time $\poly(n/\delta)$ and space $O(\log(n/\delta))$ such that with probability $1-2^{-\poly(n/\delta)}$, it outputs $w^\dagger M^i v$ with additive error $\delta$.
\end{lemma}
Note that $v_i(j)=e_j^\dagger M^i e_1$. Thus, by repeating the subroutine from \Cref{grzlemma} $\poly\pbra{\tfrac{nT}{\delta}}$ times with parameters $w=e_j,v=e_1,i$ and $\delta/n$, a quantum logspace prover can with probability at least $\tfrac{3}{4}$, estimate each $v_i(j)$ to $\delta/n$ additive accuracy for all $i\in[T]$ and $j\in[n]$. In this case, we have, $\vabs{{v'_i}-v_i}_2 \le  \vabs{{v'_i}-v_i}_\infty \cdot n \le \delta$. This completes the proof of \Cref{claimqprover}.
\end{proof}
We now complete the proof of \Cref{claimverifier}

\begin{proof}[Proof of \Cref{claimverifier}] The verifier's algorithm is formally described in \textbf{Algorithm 1}. The informal description is as follows. The verifier will try to check that $\widetilde{M}(v'_{i-1})$ is approximately equal to $v'_i$ for all $i\in[T]$. However, to do this in a streaming fashion, the verifier will instead test that a random linear combination of these approximate equations holds. To reduce the randomness from $T$ to $O(\log n)$, instead of using a truly random combination of the equations the verifier uses a pseudorandom combination drawn using a 4-wise independent collection of $\{-1,1\}$-random variables. This is similar to the $\ell_2$-frequency estimation algorithm in~\cite{ams}.

	\begin{algorithm}
		\caption{ Algorithm for Verifier in \Cref{claimverifier}}
		\SetKwInOut{Input}{Input}\SetKwInOut{Output}{Output}
		\Input{An $n\times n$ unitary matrix $M$, parameters $T\le \poly(n),\tfrac{1}{10^4 T^2}\ge \delta \ge \tfrac{1}{\poly(n)}$ and read-once access to a stream of vectors $v'_0,\ldots,v'_T\in \mathbb{R}^n$. }
		\Output{
			If the sequence is $\delta$-good for $M$, then return $\bot$ with probability at most $\tfrac{1}{4}$. 
			If $\vabs{v'_T-v_T}_2 \ge  \tfrac{1}{5}$, return $\bot$ with probability at least $\tfrac{3}{4}$.}
		
		\Begin{
			Round down each entry of the input matrix $M$ to $\frac{\delta}{6n^2T}$ additive error to produce a matrix $\widetilde{M}$ so that $\vabs{M-\widetilde{M}}_2 \le \tfrac{\delta}{6 T }$.
			
			Return $\bot$ if $v'_0\neq e_1$.\;
			
			\For{$t= 1$ \KwTo  $11$}
			{Sample $\alpha_{i,j}\in \{-1,1\}$ for $i\in [T],j\in [n]$ from a collection of 4-wise independent $\{-1,1\}$-random variables with mean 0.\;
				
				Compute $\Delta:=\sum_{i\in[T],j\in[n]} \alpha_{i,j}\cdot w_{i,j}$ where for $i\in [T],j\in [n]$, we have $w_{i,j}:= (\widetilde{M}(v'_{i-1}))(j)-v'_{i}(j)$.\;
				
				Return $\bot$ if $|\Delta| > 30  T \delta  $. 
			}
		}
	\end{algorithm}

	\paragraph*{Time \& Space Complexity of this Algorithm:} One can sample from a collection of 4-wise independent $\{-1,1\}$-random variables of size $O(nT)$ in logspace using only $O(\log(nT))$ bits of randomness~\cite{ams}. Note that the quantity $\Delta\triangleq \underset{\substack{i\in [T]\\j\in [n]}}{\sum} \alpha_{i,j}\cdot \pbra{(\widetilde{M}(v'_{i-1}))(j)-v'_{i}(j)}$ can be expressed ${\sum}_{\substack{ i\in\{0,\ldots,T\}\\j\in [n]}} \beta_{i,j} v'_i(j)$ where $\beta_{i,j}$ are coefficients that depend only on the entries of $\widetilde{M}$ and $\alpha$, and can be computed in logspace. Thus, a logspace algorithm can read the stream of $v'_i(j)$ for $i=0,\ldots,T$ and $j\in [n]$ once from left to right and compute $\Delta\triangleq \sum_{i,j}  \beta_{i,j} v'_i(j)$ in a streaming fashion. As the entries of the matrices and the vectors are $O(\log (n))$ bits long, the arithmetic can be done in logspace. The time complexity of this process is hence $\poly(n)$ and the space complexity is $O(\log( n))$.

	We now move on to the completeness and soundness. First, we make some observations.  Let $w\in \mathbb{R}^{nT}$ be defined at $i\in [T],j\in [n]$ by $w_{i,j}\triangleq (\widetilde{M}(v'_{i-1}))(j)-v'_{i}(j)$. Let $\widetilde{v}_0,\ldots,\widetilde{v}_T$ be defined by $\widetilde{v}_i=\widetilde{M}^i(e_1)$ for all $i\in[T]\cup\{0\}$. Since $\vabs{\widetilde{M}-M}_2\le \tfrac{1}{6\delta T}$ and $\|M\|_2\le 1$, 
		\begin{align}\begin{split}\label{eqqtildeM}
			\text{for all }i\in[T],\vabs{\widetilde{M}^i-M^i}_2&\le \pbra{1+\frac{\delta}{6  T}}^i -1 \le \frac{\delta}{2}.\\
	\end{split}\end{align}
	(In particular, $\vabs{\widetilde{M}^i}_2 \le 1 + \delta/2$.) Thus, 
	\begin{align}\label{eqq} \text{for all }i\in[T],\|\widetilde{v}_i-v_i\|_2 &\triangleq \vabs{ \widetilde{M}^i(e_1)-M^i(e_1)}_2\le \vabs{\widetilde{M}^i-{M}^i}_2\le  \frac{\delta}{2}.
	\end{align}

	\paragraph*{Completeness of the Algorithm:} 
	Suppose $v'_0,\ldots,v'_T$ is a $\delta$-good sequence, then $\|v'_i-v_i\|_2 \le \delta$ for all $i\in [T]$ and $v'_0=e_1$. Since $M$ is a contraction map with respect to $\| \cdot \|_2$, this along with \Cref{eqqtildeM} implies that for all $i\in[T]$, \begin{align*} \vabs{\widetilde{M}(v'_{i-1})-v'_{i}}_2&\le  \vabs{\widetilde{M}(v'_{i-1})-M(v'_{i-1})}_2+ \vabs{ M(v'_{i-1})-M(v_{i-1})}_2\\
		&+\vabs{M(v_{i-1})-v_{i}}_2 + \|v_{i}-v'_{i}\|_2\\
		&\le \vabs{\widetilde{M}-M}_2\cdot \|v'_{i-1}\|_2+ \|v_{i-1}-v'_{i-1}\|_2 +\|v_{i}-{v}'_{i}\|_2\\
		&\le \tfrac{ \delta}{6T}\cdot (1+\delta )+ \delta+ \delta  \le 3\delta  . \end{align*}
	Thus, $\|w\|_2 \le 3T \delta$. Consider the quantity $\langle \alpha,w\rangle=\sum_{i,j} \alpha_{i,j} w_{i,j}$ that the algorithm estimates. Note that $\mathbb{E}\sbra{\langle \alpha,w\rangle}=0$ and that $\mathbb{E}\sbra{\langle \alpha,w\rangle^2}=\sum_{i,j} w_{i,j}^2 $. Chebyshev's Inequality implies that with probability at least $0.99$, we have $|\langle \alpha,w\rangle|\le  30 T\delta  $. This implies that with probability at least $(0.99)^{11}\ge 0.8$, every iteration of the inner loop in {\bf Algorithm 1} does not reject. 
	
	\paragraph*{Soundness of the Algorithm:} Suppose a dishonest prover produces a stream $v'_0,\ldots,v'_T$ such that $\vabs{v'_T-v_T}_2\ge \tfrac{1}{5}$. 	The verifier always returns $\bot$ if $v'_0\neq e_1$, so we may assume that $v'_0=e_1$. Let $\eps=\tfrac{1}{20T}$. We argue that for some $i\in [T]$, we must have $\vabs{w_{i}}_2\ge \eps$. Assume by contradiction that $\vabs{\widetilde{M}(v'_{i-1})-v'_i}_2 \le \eps$  for all $i\in [T]$. Hence, by Triangle Inequality and \Cref{eqqtildeM}, (and since $\widetilde{v}_0=e_1$) we have
	\begin{align*}\begin{split}
			\vabs{\widetilde{v}_T-v'_T}_2=\vabs{\widetilde{M}^T(v'_0)-v'_T}_2 &\le \sum_i \vabs{\widetilde{M}^{T-(i-1)}(v'_{i-1})-\widetilde{M}^{T-i}(v'_i)}_2 \\
			&\le \sum_i \vabs{\widetilde{M}^{T-i}}_2 \cdot \vabs{ \widetilde{M}(v'_{i-1})-v'_i}_2\\
			&\le \sum_i \pbra{1+\tfrac{\delta}{2}} \cdot \eps\\
			& \le 2T\eps.
	\end{split}\end{align*} 
	\Cref{eqq} implies that $\vabs{ \widetilde{v}_T  - v_T}_2 \le \tfrac{\delta}{2}$. This implies that $\vabs{v'_T-v_T}_2 \le \tfrac{\delta}{2}+2T \eps$. We assumed that  $ \vabs{v_T-v'_T}_2 \ge \tfrac{1}{5}$. Hence, it follows that
	\[ \tfrac{1}{5}\le \tfrac{\delta}{2}+2T \eps. \] 
	Since we chose $\eps=\tfrac{1}{20T}$ and $\delta\le 1/10$, this is a contradiction. Thus, we must have $\|w\|_2\ge \eps$. Note that $\mathbb{E}\sbra{\langle \alpha,w\rangle}=0$ and $\mathbb{E}\sbra{\langle \alpha,w\rangle^2}=\|w\|_2^2$. Furthermore, 
	\[ \mathbb{E}\sbra{\langle \alpha,w\rangle^4}=\mathbb{E}\sbra{\sum_{i,j,k,l} w_iw_jw_kw_l\alpha_i \alpha_j\alpha_k \alpha_l } \le 6 \sum_{i,j}  w_i^2w_j^2 \le 6\|w\|_2^4 \]
	Here, we used the fact that the random variables are 4-wise independent. The Paley-Zygmund Inequality implies that 
	\[ \p\sbra{ \langle \alpha, w\rangle^2 \ge \frac{1}{10}\cdot  \|w\|_2^2 } \ge \pbra{1-\frac{1}{10}}^2\cdot \frac{\pbra{\mathbb{E}\sbra{\langle \alpha, w\rangle^2}}^2}{\mathbb{E}\sbra{\langle \alpha, w\rangle^4}}\ge\frac{1}{8}. \]
	This, along with the fact that $ \|w\|_2\ge \eps$ implies that $\p\sbra{\abs{ \langle \alpha, w\rangle} \ge  \frac{\eps}{10}  } \ge \frac{1}{8}$. By repeating this experiment $11$ times, we can ensure that with probability at least $1-(1-1/8)^{11}\ge 3/4$, we find at least one instance so that $\abs{\langle \alpha, w\rangle }\ge \tfrac{\eps}{10}$. Since $\delta \le \tfrac{1}{10^4 T^2}$ and $\eps = \tfrac{1}{20T}$, we have 
	\[ \tfrac{\eps}{10} > 30  T\delta \]
	Thus, with probability at least $3/4$, we have $\abs{\langle \alpha,w\rangle}> 30 T\delta$. This implies that the algorithm returns $\bot$ with probability at least $3/4$. 
	
\end{proof}

\bibliographystyle{alpha}
\bibliography{ref}

\section{Appendix} 
\subsection{A $\BQL$-complete Problem}

We prove \Cref{bqlcomplete} which states that the Unitary Matrix Powering Problem is complete for $\BQL$. As before, it suffices to reduce all $\BQL$ problems to this problem.

Consider any $\mathcal{F}=\{f_n:\{0,1\}^n\to \{0,1\}\}_{n\in \mathbb{N}}$ in $\BQL.$ As per the definition\footnote{Strictly speaking, this definition is for a unitary variant of $\BQL$, however, in~\cite{fr20} it is shown that all problems in $\BQL$ are solvable by this unitary variant.} in~\cite{fr20}, this means that there exists a logspace-uniform family of quantum circuits $\{Q_n(x)\}_{n\in \Nbb}$, consisting of only unitary operators  where $Q_n(x)$ acts on $m=O(\log n)$ qubits with the following property. If the initial state is $\ket{0^m}$ and the first qubit of the final state is measured, then $f_n(x)=1$ if the outcome is 0 with probability at least $4/5$ and $f_n(x)=0$ if the outcome is 0 with probability at most $1/5$. Let $T_n(x)$ be the number of operators of the quantum circuit $Q_n(x)$ and $m$ be the number of qubits. Define a unitary matrix $U_n(x)$ in $(T_n(x)+1)\times 2^{m}$ dimensions as follows.  We first partition the rows and columns of $U_n(x)$ into $T_n(x)+1$ parts based on the value of the first $\log(\lceil T_n(x)+1\rceil)$ coordinates. For all $i\in [T_n(x)]$, define the $(i+1,i)$-th block of $U_n(x)$ to be the $i$-th operator in the circuit $Q_n(x)$. Define the $(1,T_n(x)+1)$-th block of $U_n(x)$ to be the identity matrix. All other blocks of $U_n(x)$ are defined to be zero. Since $T_n(x)\le \poly(n)$ and $m\le O(\log n)$, this is a unitary operator in $\poly(n)$ dimensions. Let $\Pi_n(x)$ be a projection matrix in $(T_n(x)+1)\times 2^{m}$ dimensions that projects onto the basis states $\{\ket{i,j}\mid i=T_n(x)+1, j\in [2^m],j_1=0\}$. 

Firstly, each entry of the unitary matrix $U_n(x)$ and the projection matrix $\Pi_n(x)$ can be computed by a deterministic logspace algorithm.  Observe that the vector $U^i_n(x)(e_1)$ is supported only on coordinates in $\{i+1\}\times [2^m]$, furthermore, when restricted to these coordinates, this vector precisely captures the state of the qubits in $Q_n(x)$ after applying the first $i$ operators. It follows that the probability that the circuit $Q_n(x)$ outputs 1 is precisely $\|\Pi U^{T_n(x)}(e_1)\|^2$. Thus, given any $x\in \supp(f_n)$, we can produce in deterministic logspace, a unitary matrix $U_n(x)$ and a projection matrix $\Pi_n(x)$ in $\poly(n)$ dimensions and a parameter $T\le \poly(n)$ such that $f_n(x)=1$ if $\|\Pi U^{T_n(x)}(e_1)\|^2\ge 4/5$ and $f_n(x)=0$ if $\|\Pi U^{T_n(x)}(e_1)\|^2\le 1/5$. This shows that the Unitary Matrix Powering Problem is complete for $\BQL$.
\end{document}